\documentclass[twocolumn,amsmath,amssymb]{revtex4}

\usepackage{graphicx}
\usepackage{dcolumn}
\usepackage{bm}

\begin{document}
\title{Comment on ``Nernst effect in poor conductors and in the cuprate superconductors''}

\date{\today}
\maketitle

In a recent letter, Alexandrov and Zavaritsky(AZ)\cite{alexandrov}
propose a model to explain the anomalous Nernst signal observed in
the normal state of hole-doped cuprates\cite{xu,wang, capan}
without invoking superconducting fluctuations. In the proposed
picture\cite{alexandrov},  the thermomagnetic contributions from
itinerant and localized carriers ``interfere'' to produce a large
Nernst signal. The authors argue that this picture is relevant to
the high-T$_{c}$ cuprates and can quantitatively describe the
experimental data. The aim of this comment is to recall that one
of the main assumptions of the authors regarding the experimental
data is false.  Contrary to what they assume, at low temperature
and high magnetic fields, the magnitude of the Nernst signal, ($N$
or $e_y$ ) easily exceeds the product of the Seebeck coefficient,
$S$, and the Hall angle, $\tan\theta$.

Comparing the magnitude of the Nernst signal with $S\tan\theta$ is
instructive to identify the origin of the anomalous Nernst signal in
the normal state. The transverse voltage created by a longitudinal
thermal gradient, $e_{y}=\frac{E_{y}}{\nabla_{x}T}$ is expressed as
the difference of two terms, $\rho\alpha_{xy}$ ($\rho$ is
resistivity and $\alpha_{xy}$ represents the off-diagonal Peltier
conductivity) and $S\tan\theta$\cite{wang}. These two terms cancel
out in a simple one-band metal\cite{wang} but not in metals with
different type of carriers where the two terms can add up\cite{bel}.
In the scenario proposed by AZ, there is no cancelation either and
the contribution of localized and itinerant carriers add up to
produce a large Nernst signal. Namely,  while localized carriers
provide a large thermopower, only itinerant carriers contribute to
the Hall conductivity, $\sigma_{xy}$. As a consequence, a
significant Nernst signal is found, comparable in magnitude to
$S\tan\theta$ (and $\rho\alpha_{xy}$). In their account of the
experimental data, they claim that ``... $e_y$ and $S\tan\theta$ are
of the same order at sufficiently low
temperatures.''\cite{alexandrov}

Let us examine this statement, which is crucial for the relevance
of the model since they write :``If carriers are fermions,
S$tan\Theta_{H}$ should be larger than or of the same order of
magnitude as e$_y$, because their ratio is proportional to
$\sigma_{xx}/\sigma_{l}>>1$ in our model.''\cite{alexandrov} Fig.
1 presents the case of La$_{1.94}$Sr$_{0.06}$CuO$_{4}$ which lies
close to the superconductor-insulator boundary. As seen in the
figure, the application of a magnetic field of 12T leads to the
emergence of the well-known non-metallic resistivity of the
underdoped cuprates\cite{ando}. In spite of the apparent
destruction of superconductivity, a finite positive Nernst signal
survives at this magnetic field in the same sample. Were it due to
the presence of localized carriers, then, in the AZ picture, one
would expect $e_y \sim S \tan \theta$. However, as seen in the
right panel, the Nernst signal becomes almost an order of
magnitude larger than $S\tan\theta$ (measured in the same magnetic
field and at the same temperature). To the best of our knowledge,
this observation (i.e. $e_y \gg S \tan \theta$) holds in all those
cases for which both Nernst data and resistive evidence for
field-induced localization are available (i.e. LSCO and Bi-2201 at
sufficiently low temperature, high magnetic field and low doping
level)\cite{capan}. We note also that  at sufficiently large
magnetic fields, the positive Nernst signal fades away while the
localization does not\cite{wang}.  Moreover, at lower doping
levels (i.e. $x<0.05$), LSCO displays localized behavior even in
zero magnetic field but no positive $e_y$\cite{wang}. These
observations do not seem compatible with the idea of localized
excitations as the central source of the positive $e_y$.
\begin{figure}
\resizebox{!}{0.35\textwidth}{\includegraphics{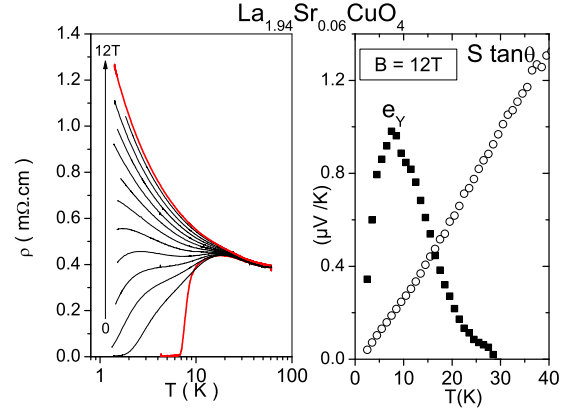}}
\caption{\label{fig1}Left: Emergence of a field-induced
``insulating'' behavior in underdoped LSCO as revealed by
temperature-dependence of resistivity. Right: The Nernst signal
and S$\tan\theta$ at 12T as a function of temperature for the same
sample.}
\end{figure}

Any proposed alternative to the superconducting
fluctuations\cite{xu,wang,capan} as the origin of the anomalous
Nernst signal is expected to explain these features. The proposed
model\cite{alexandrov} does not meet this requirement to qualify
as a relevant explanation of the anomalous Nernst signal in
cuprates.\\
\textbf{Cigdem Capan$^{1}$} and \textbf{Kamran Behnia$^{2}$}\\
\emph{(1)Department of Physics and Astronomy, Louisiana State University, Baton Rouge, Louisiana 70803\\
(2)Laboratoire de Physique Quantique(CNRS), ESPCI, 10 Rue de
Vauquelin, 75005 Paris, France}

\end{document}